\documentclass{JAC2003}
\usepackage{graphicx}
\def\beq{\begin{equation}}   
\def\eeq{\end{equation}} 
\begin{document}
\title{MAPS FOR ELECTERON CLOUDS:\\ APPLICATION TO LHC CONDITIONING}
\author{T. Demma, R. Cimino, A. Drago, INFN-LNF, Frascati (Italy),\\
S. Petracca, A. Stabile, University of Sannio, Benevento (Italy).}
\maketitle

\begin{abstract}
In this communication we present a generalization of the map formalism, introduced in\cite{Iriso} and \cite{noi}, to the analysis of electron flux at the chamber wall with particular reference to the exploration of LHC conditioning scenarios.
\end{abstract}

\section{INTRODUCTION}
The electron cloud driven effects can limit the ability of recently build or planned accelerators to reach their design parameters. The secondary emission yield reduction (called "scrubbing") due to the fact that the electrons of the cloud hit the vacuum chamber wall, modifying its surface properties, may minimize any disturbing effects of the cloud to the beam. Surface scrubbing was studied in various experiments, by measuring the electron dose (the number of impinging electrons per unit area on sample surfaces) dependence of SEY yield. All the available experiments found in literature have been performed by bombarding technological metal surfaces with electron beams of fixed energy as $300-500 eV$ and $2.5 keV$. They showed that even a low electron exposure of about $10^{-6} Cmm^{-2}$ the SEY of material start to decrease, reaching its lower value after about $10^{-2} C/mm^{-2}$. Altough these investigations gave informations about conditioning in accelerators, they are not complete and other studies are required to clarify the scrubbing dependence on the bombarding dose of impinging electron beams, since this parameter is missing. The dependence of "scrubbing" efficiency on beam and chamber parameters can be deduced from e-cloud simulation codes (e.g. PEI \cite{pei} , POSINST \cite{posinst}, and ECLOUD \cite{ecloud}) modeling the involved physics in full detail. 
In \cite{Iriso} it was  shown that the evolution of the electron cloud density from one bunch passage to the next can be described using a cubic map whose parameters can be extrapolated from simulations, and are functions of the beam parameters and of the beam pipe features. Simulations based on the above map are orders of magnitude
faster than those based on particle-tracking codes. In this paper we generalize the map formalism, introduced in\cite{Iriso} and \cite{noi}, to the analysis of electron flux at the chamber wall.
 
\section{Map Formalism}

\begin{figure}[htb]
\centering
\includegraphics*[width=80mm]{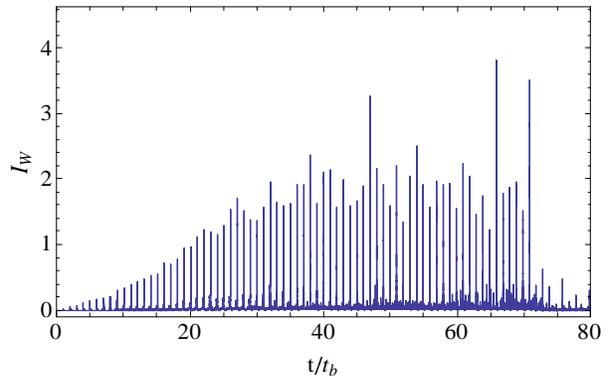}
\caption{instantaneous current $I_W$ of electrons bombarding the wall of an LHC arc dipole computed with ECLOUD. The case shown corresponds to a filling pattern featuring 72 charged bunches, with bunch charge of $N=1.8\cdot10^{11}$ protons, followed by 8 empty (zero-charge) bunches. The assumed bunch spacing is 7.48 m, and the SEY is $\delta_{max}=1.8$.}
\label{fig:fig1}
\end{figure}

The time evolution of the instantaneous current $I_W$
of electrons bombarding the wall of an LHC arc dipole, computed by
ECLOUD, is shown in Fig. \ref{fig:fig1}, for a filling pattern consisting
of a train of 72 bunches followed by gaps of 8 empty (zero
charge) bunches, and the beam/pipe parameters collected
in Table I.
\begin{table}[hbt]	
\begin{center}	
\caption{Parameters used for ECLOUD simulations.}	
\begin{tabular}{lcc}                                                  		
\hline
\hline		
parameter & units & value \\		
\hline		
beam particle energy & $GeV$ & $7000$\\		
bunch spacing $t_b$ & $ns$ & $25$\\		
bunch length & $m$ & $0.075$\\		
number of  bunches $N_b$ &-& $72$\\
bunch gap $N_g$ &-& $8$\\		
no. of particles per bunch & -- & $1.2\cdot10^{11}$ \\		
bending field $B$ & $T$ & $8.4$\\		
length of bending magnet & $m$ & $1$\\		
vacuum screen half height & $m$ & $0.018$\\		
vacuum screen half width & $m$ & $0.022$\\ 		
circumference & $m$ & 27000\\		
primary photo-emission yield &-& $7.98\cdot10^{-4}$\\    
maximum $SEY$ $\delta_{max}$ &-& $1.5$\\		
energy for max. $SEY$ $E_{max}$ & eV & 237.125\\		
\hline
\hline	
\end{tabular}
\label{tab1}
\end{center}
\end{table}
\begin{figure}[htb]
\centering
\includegraphics*[width=80mm]{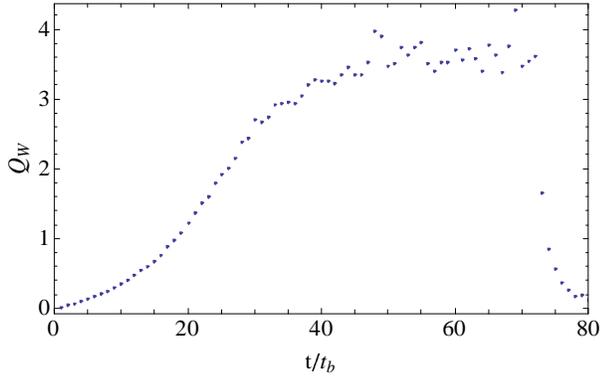}
\caption{"bunch-by-bunch" electron dose $Q_m$. The case shown corresponds to a filling pattern featuring 72 charged bunches, with bunch charge of $N=1.8\cdot10^{11}$ protons, followed by 8 empty (zero-charge) bunches. The assumed bunch spacing is 7.48 m, and the SEY is $\delta_{max}=1.8$.}
\label{fig:fig2}
\end{figure}
The "bunch-by-bunch" electron dose $Q_n$ given to the chamber walls is obtained by integrating the current $I_W$ in the intervals between successive bunch passages, and is showed in Fig. \ref{fig:fig2}. The electron dose grows exponentially in time as more and more bunches passes by, until saturation occurs. The subsequent decay corresponds to the successive passage of the empty bunch train.
\begin{figure}[htb]
\centering
\includegraphics*[width=80mm]{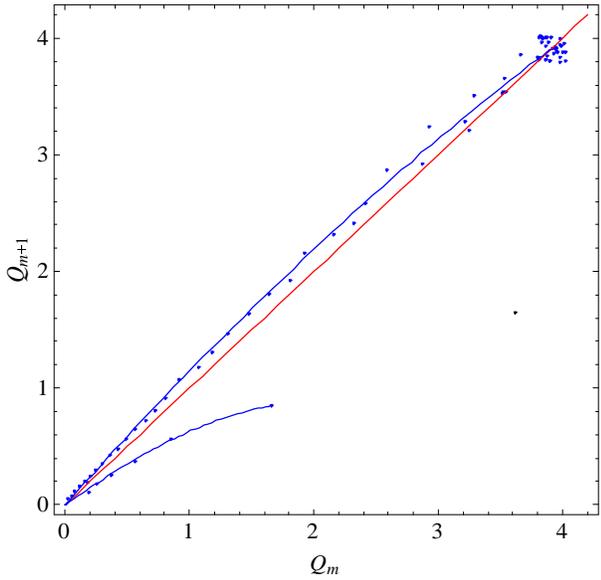}
\caption{"bunch-by-bunch" electron dose map $Q_{m+1}$ vs $Q_m$. Circle
markers: ECLOUD simulations ($\delta_{max}=1.8$, all other parameters
as in Table I). The red line represents saturation ($\rho_{m+1}=\rho_m$). Markers above the saturation line describe the buildup. Markers below the saturation line describe the e-cloud decay.
The blue lines are the corresponding cubic fits.
Transitions between filled and empty bunch trains are shown
as black circles.}
\label{fig:fig3}
\end{figure}
\begin{figure}[htb]
\centering
\includegraphics*[width=80mm]{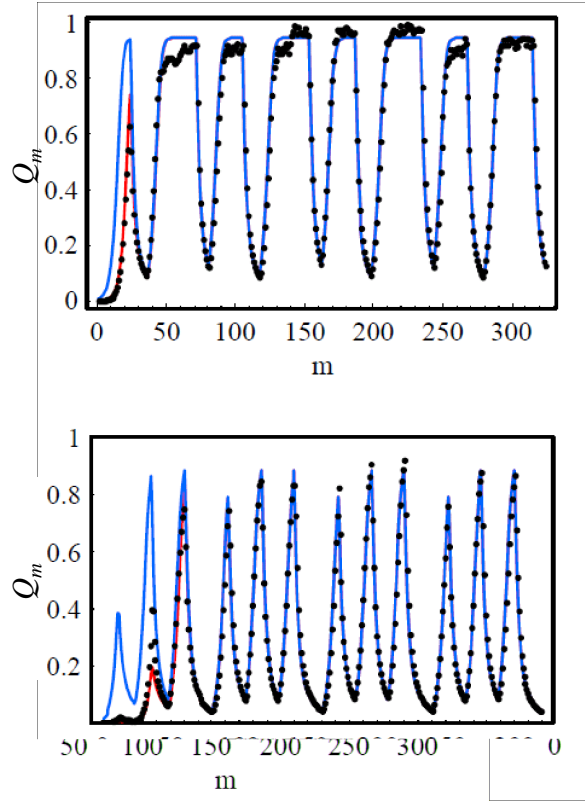}
\caption{Electron dose "bunch-by-bunch" evolution for two
different bunch-train filling patterns, for $N=1.8\cdot10^{11}$ and
$\delta_{max}=1.8$. Top: (24f,12e,36f); bottom:
$3\times(12f 12e)$. In both cases, successive bunch trains are
separated by gaps whose length corresponds to 28 (empty)
bunches. Black markers: ECLOUD results. Blue and red lines:
map results corresponding to different initial electron doses.}
\label{fig:fig4}
\end{figure}
Figure \ref{fig:fig3} shows the behavior of the electron dose (computed as explained above) after the passage of
bunch $m$, denoted as $Q_{m+1}$, as a function of $Q_m$. The red line in Fig. \ref{fig:fig3} corresponds
to saturation (fixed points of the $Q_m \rightarrow Q_{m+1}$ map).
As more and more bunches pass by, the initially small
electron-cloud density builds up (points above the red line,
$Q_m > Q_{m+1}$), eventually approaching saturation. In the
saturation regime, the $Q_m$ tend to cluster along the red
line. Points below the red line ($Q_m < Q_{m+1}$) describe the
decay regime. The continuous curves in Fig. \ref{fig:fig3} correspond to homogeneous
cubic fits,
\beq
Q_{m+1}=a Q_m + b Q_m + c Q_m
\label{eq:map}
\eeq
which are seen to reproduce the data quite well. The map
idea introduced in \cite{Iriso} and \cite{noi} for the e-cloud density thus works
also for the "bunch-by-bunch" electron dose.
The only exceptions are represented by the transitions
between filled and empty bunch subtrains, represented by
the square markers in Fig. \ref{fig:fig2}. This is not unexpected, and
was already noted in \cite{Iriso}.
The three terms in the map (1) describe, respectively, the
exponential growth/decay mechanism (linear term, larger/
smaller than 1, respectively), the space charge effects
leading to saturation (quadratic term, whose sign reflects
the concavity of the curves), and an additional correction
(cubic term) embodying small corrections.
At present, there is only partial clear physical insight
into the dependence of the above map parameters on the
problem's (beam and pipe) configuration, and their values
must be deduced empirically from numerical simulations.
Once the coefficients have been determined, however,
the model is accurate for all filling patterns, as further
illustrated in Fig. \ref{fig:fig4} where are compared results 
obtained by ECLOUD and the cubic map formalism using
the map coefficients corresponding to the reference filling
pattern of LHC (72 charged bunches) to predict the electron dose "bunch-by-bunch" evolution for different filling patterns.
In particular, regardless of the initial longitudinal
electron density, the map results agree within an error
range of $10\%$ for all bunch filling patterns.

\section{CONCLUSIONS AND OUTLOOK}

The "bunch-by-bunch" electron dose delivered to LHC dipoles chambers can be
described by a cubic map. The coefficients of this map
depend on the pipe and beam parameters, and can be
simply deduced from e-cloud simulation codes modeling
the involved physics in full detail. Remarkably, if all other
parameters (namely, the bunch charge N, the SEY, and the
pipe parameters) are held fixed, the map coefficients basically
do not depend on the filling pattern.
The map can be thus used as a quick and (not so) dirty
tool for finding filling patterns yielding the highest "scrubbing" efficiency compatible with a given set of beam parameters.
Unfortunately, at present, no physical model for relating all
map coefficients to the problem's parameter is still available. We are working
toward adapting approach presented in \cite{linterm1} and \cite{linterm2}, for the computation of the linear
coefficient in the map describing the evolution of the electron cloud density, to the case presented here.

\end{document}